\documentclass[%
 aip,
 amsmath,amssymb,
 reprint,%
]{revtex4-1}

\usepackage{graphicx}% Include figure files
\usepackage{dcolumn}% Align table columns on decimal point
\usepackage{bm}% bold math
\usepackage{siunitx}
\usepackage[utf8]{inputenc}
\usepackage[T1]{fontenc}
\usepackage{mathptmx}

\begin{document}

\preprint{AIP/123-QED}

\title[Plasma formation and relaxation dynamics in fused silica driven by femtosecond short-wavelength infrared laser pulses]{Plasma formation and relaxation dynamics in fused silica driven by femtosecond short-wavelength infrared laser pulses}
% Force line breaks with \\

\author{P. Jürgens}
\email{juergens@mbi-berlin.de}
\affiliation{Max-Born-Institute for Nonlinear Optics and Short Pulse Spectroscopy, Max-Born-Str. 2A, D-12489 Berlin, Germany}
\author{M. J. J. Vrakking}%
\affiliation{Max-Born-Institute for Nonlinear Optics and Short Pulse Spectroscopy, Max-Born-Str. 2A, D-12489 Berlin, Germany}
%\\This line break forced with \textbackslash\textbackslash
%

\author{R. Stoian}
\affiliation{Laboratoire Hubert Curien, UMR CNRS 5516, Unversit{\'e} de Lyon, Unversit{\'e} Jean Monnet, 42000 Saint Etienne, France}

\author{A. Mermillod-Blondin}
\affiliation{Max-Born-Institute for Nonlinear Optics and Short Pulse Spectroscopy, Max-Born-Str. 2A, D-12489 Berlin, Germany}

\date{\today}% It is always \today, today,

\begin{abstract}
Laser-induced plasma formation and subsequent relaxation in solid dielectrics is the precursor to structural modifications that are accompanied by a permanent alteration of material properties. The decay of the electron-hole plasma through distinct relaxation channels determines the properties of the resulting modification. Based on an experimental arrangement combining a time-resolved transmission measurement with a cross-phase modulation measurement, we isolate the plasma formation and relaxation dynamics in the bulk of amorphous fused silica excited by femtosecond short-wavelength infrared ($\lambda =\,$\SI{2100}{\nano\meter}) laser pulses. Whereas the relaxation time of the generated electron-hole plasma was so far assumed to be constant, our findings indicate an intensity-dependent relaxation time. We attribute this intensity dependence to vibrational detrapping of self-trapped excitons.
\end{abstract}

\maketitle

\begin{figure}[htbp]
\centering
\fbox{\includegraphics[width=\linewidth]{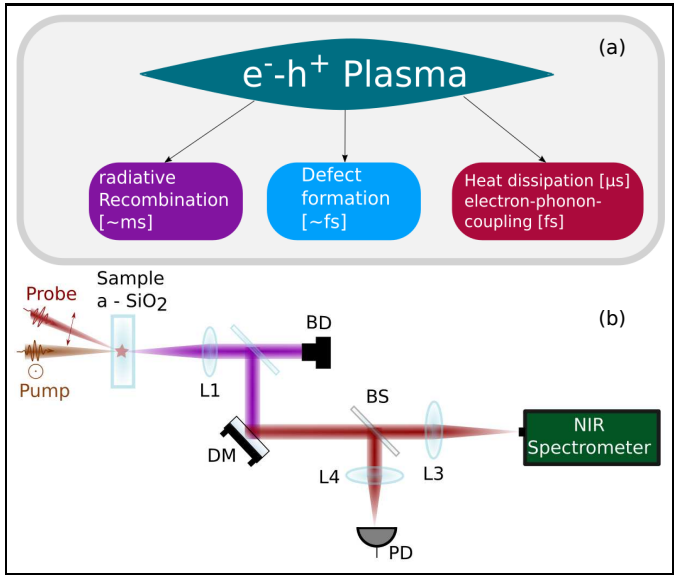}}
\caption{(a) The three relaxation mechanisms for an electron-hole plasma created by intense laser irradiation of a solid dielectric. (b) Experimental setup for the cross-polarized, close-to-collinear pump-probe investigations. L1 - L4: Lenses, BD: Beam dump, DM: Dielectric mirror (HR \SI{800}{\nano\meter}), BS: Beam splitter, PD: Photodiode. NIR Spectrometer: Near-infrared spectrometer.}
\label{fig:setup}
\end{figure}

\begin{figure}[htbp]
\centering
\fbox{\includegraphics[width=\linewidth]{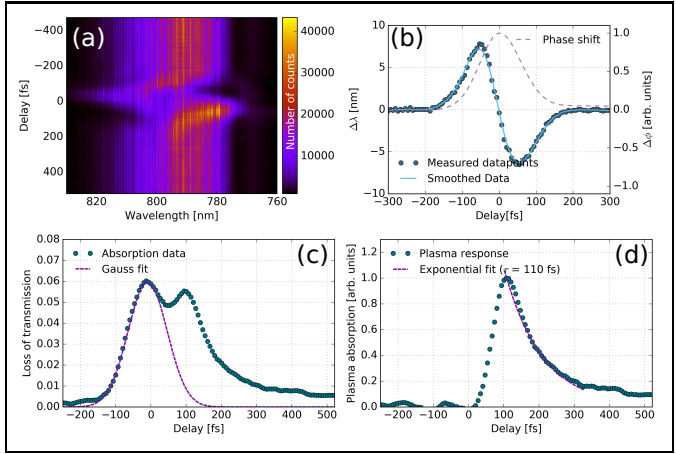}}
\caption{(a) + (b) Experimental results obtained at a pump laser intensity of \SI{11}{\tera\watt\per\centi\meter\squared}. (a) Time-frequency map of the probe spectrum as a function of the delay between pump and probe laser pulse, revealing a delay-dependent shift of the probe spectrum due to cross-phase modulation (XPM). (b) Extracted dynamics of the central probe wavelength together with the phase shift induced by the SWIR pump laser pulse in the fused silica target. (c) + (d) Experimental absorption measurements obtained using the photodiode, recorded for the same incident pump laser intensity. (c) Total loss of transmission as a function of the delay between the pump and the probe laser pulse. (d) Isolated plasma dynamics with an exponential fit to the decay of the absorption profile.}
\label{fig:results_1}
\end{figure}

\begin{figure}[htbp]
\centering
\fbox{\includegraphics[width=\linewidth]{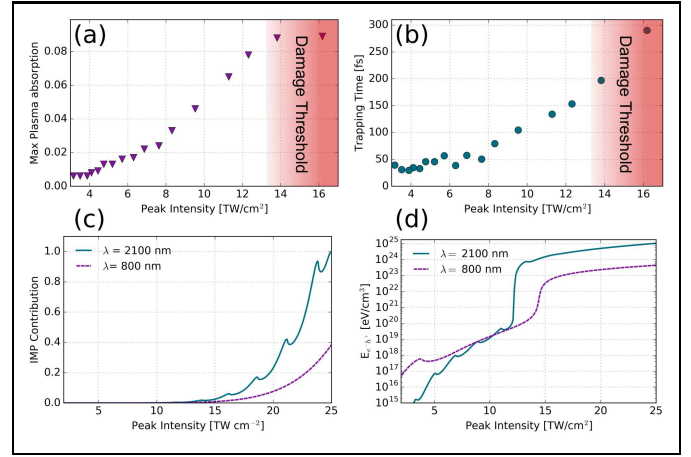}}
\caption{(a) + (b) Experimental determination of the maximum absorption strength and the relaxation times obtained from the pump intensity scans. (a) Maximum value of the retrieved plasma absorption as a function of the excitation strength. (b) Intensity dependence of the relaxation times obtained by applying exponential fits to the decay of the isolated plasma absorption profiles. (c) Numerical simulation according to Eq.~\ref{eq:efcb} of the energy coupled into the free carrier bath as a function of the laser intensity for a central wavelength of $2100$ and \SI{800}{\nano\meter}. (d) Numerical results for the relative contribution of IMP with respect to SFI obtained by solving Eq.~\ref{eq:sre}. The pulse duration in (c) and (d) corresponds to 20 optical cycles of the driving field.}
\label{fig:results_3}
\end{figure}

The formation of an electron-hole plasma in solid dielectrics under the influence of intense femtosecond laser pulses holds the key to a wide range of subsequent processes that are relevant to both fundamental physics and industrial applications. The availability of reliable femtosecond laser sources satisfying industrial standards has led to the establishment of these systems in the field of precise bulk micromachining and laser ablation \cite{Gattass2008}. Moreover, as the main limiting factor for stability and output power of high-power laser systems, ultrashort pulse laser-induced damage in optical components has been proven to be based on electronic processes \cite{Stuart1995, Jupe2009} and is strongly connected to the surpassing of a critical plasma density \cite{Sudrie2002}. The electron-hole plasma is mainly generated by two competing ionization mechanisms. Whereas free carriers are generated by laser-induced strong-field ionization (SFI), already excited conduction band electrons can excite further electrons from the valence band by electron-impact ionization (IMP). These two excitation channels are the main mechanisms of energy coupling into the material. The subsequent relaxation of the quasi-free carriers determines the reorganization of the material and therefore the resulting modification of the target. Three different mechanisms are usually associated with the decay of the conduction band electron density [see Fig.~\ref{fig:setup}(a)]. While direct radiative recombination generally occurs on a ms timescale, defect formation and eletron-phonon coupling occur on timescales of $\approx$ \SI{100}{\femto\second} \cite{Gamaly2013}. The influence of heat is often considered as too slow to influence ultrafast trapping phenomena since heat dissipation occurs on a microsecond timescale. 
In strongly coupled dielectrics the ultrafast formation of defects such as non-bridging oxygen-hole centers and E' centers is preceded by self-trapping of quasi-free carriers \cite{Itoh1994}. The self-trapping is associated with the matrix polarizability and the ability of the carrier to exercise an influence on the molecular dipole. This interaction can be visualized as a deformation potential induced by the carrier itself in opposition to existing traps or distortions \cite{Williams1990}.  \\
To investigate the dynamics of the ultrafast relaxation processes, time-resolved studies of excitation and relaxation dynamics have been carried out extensively in the past decades \cite{Audebert1994, Martin1997, Mero2003, Sun2005, Grojo2010, Lechuga2017, Winkler2017, Ye2018, Pan2018}. The formation of self-trapped excitons (STEs) has been identified as a major source for the ultrafast depletion of the conduction band electron density in several materials including fused silica \cite{Audebert1994, Martin1997}. \\
\begin{table}[htbp]
\centering
\caption{Overview of experimentally determined relaxation times ($\tau_{\textrm{rel}}$) in fused silica from pump-probe measurements.}
\begin{tabular}{c|c|c}
\hline
Reference & Exc. wavelength & $\tau_{\textrm{rel}}$  \\
\hline
Audebert et al. \cite{Audebert1994} & \SI{620}{\nano\meter} & \SI{150}{\femto\second} \\
Martin et al. \cite{Martin1997} & \SI{395}{\nano\meter} & \SI{150}{\femto\second} \\
Li et al. \cite{Li1999} & \SI{800}{\nano\meter} & \SI{60}{\femto\second} \\
Mero et al. \cite{Mero2003} & \SI{800}{\nano\meter} & \SI{220}{\femto\second} \\
Sun et al. \cite{Sun2005} & \SI{800}{\nano\meter} & \SI{170}{\femto\second} \\
Grojo et al. \cite{Grojo2010} & \SI{800}{\nano\meter} & \SI{179}{\femto\second} \\
Pan et al. \cite{Pan2018} & \SI{800}{\nano\meter} & \SI{50}{\femto\second} \\
\hline
\end{tabular}
  \label{tab:relaxation_times}
\end{table}
As summarized in Tab.~\ref{tab:relaxation_times} a wide range of values between $50$ and \SI{220}{\femto\second} has been reported for the relaxation time of the electron-hole plasma in amorphous fused silica.
While most of these experiments have been performed in the near-infrared (NIR) \cite{Mero2003, Sun2005, Grojo2010, Pan2018, Li1999} and ultraviolet (UV) \cite{Martin1997} spectral domains, only very few experimental data exist for the interaction of short-wavelength infrared (SWIR) and mid-infrared (mid-IR) laser pulses with solid dielectrics \cite{Simanovskii2003, Grojo2013, Austin2018}. Since ultrashort pulsed long wavelength sources become more routinely available and are especially promising for high-order harmonic generation in solids \cite{Ghimire2010, Vampa2015} and precise micromachining of solid targets \cite{Grojo2013}, we present in this letter time-resolved studies on plasma formation and relaxation dynamics in amorphous fused silica driven by \SI{2.1}{\micro\meter} SWIR laser pulses.
The experimental results imply an intensity-dependent relaxation time in the bulk of fused silica that we assign to vibrational detrapping of weakly bound carriers. Our interpretation is supported by numerical simulations of the energy coupled from the laser field into the material and of the relative contribution of SFI and IMP.\\
The experimental setup [see Fig.~\ref{fig:setup}(b)] is based on a commercial Ti:sapphire laser system (Spectra Physics Spitfire Pro) delivering linearly polarized laser pulses of $\approx\,$\SI{40}{\femto\second} full width half maximum (FWHM) duration at \SI{1}{\kilo\hertz} repetition rate around a central wavelength of \SI{794}{\nano\meter}. A fraction of the pulses ($\approx$\SI{1.1}{\milli\joule}) is used to pump a commercial optical parametric amplifier (Light Conversion TOPAS C) producing an idler at a central wavelength of \SI{2100}{\nano\meter} with up to \SI{50}{\micro\joule} energy per pulse and a FWHM duration of $\approx\,$\SI{140}{\femto\second}. The idler is employed as an excitation pulse in the experiment while a small fraction of the remaining NIR laser pulse is utilized as a weak probe pulse. Both pulses are temporally and spatially overlapped inside a fused silica sample of \SI{500}{\micro\meter} thickness (UV grade fused silica Corning7980) using a cross-polarized, close-to-collinear ($\alpha_{\textrm{exc-pr}} <\,$\SI{5}{\degree}) pump probe geometry. The SWIR pump laser pulse is focused using a gold-coated off-axis parabolic mirror (OAP) with a focal length of \SI{50}{\milli\meter}, inducing a focal spot size of \SI{33}{\micro\meter} (radius at $I_0 / e^2$, measured with a knife edge technique in air) while the weak NIR probe laser pulse is focused by a silver coated focusing mirror (focal length: \SI{300}{\milli\meter}) through a central hole in the OAP, down to a spot size of \SI{75}{\micro\meter}. After recollimation the weak probe laser pulse is spectrally separated and split into two equally intense parts. One part is directed towards a photodiode (Ophir PD-10) in order to characterize the transmission of the time-delayed probe laser pulse following the modification of the target by the pump laser, while the second part is spectrally analyzed using a NIR-spectrometer (Avantes Avaspec-LS20148). \\
The fused silica sample is mounted on a computer-controlled x-y translation stage enabling scanning of the target during irradiation in order to provide a fresh spot for every laser shot. A motorized linear translation stage inserted into the beam path of the probe laser pulse allows the acquisition of transmission data and probe spectra as a function of the pump-probe delay $\tau$. Figure \ref{fig:results_1}(a) displays the detected time-resolved probe spectrum at an incident SWIR intensity of \SI{11}{\tera\watt\per\centi\meter\squared}. Cross-phase modulation (XPM) between the strong SWIR pump laser pulse and the weak NIR probe laser pulse results in strong distortions of the probe spectrum during pump-probe overlap, and is used to calibrate the zero of the time delay between the pump and the probe in the experiment. From the time-frequency map shown in Fig.~\ref{fig:results_1}(a) we extract the evolution of the central probe wavelength by calculating the center of gravity of the spectrum at each delay. The resulting variation of the central probe wavelength as a function of the pump-probe delay is shown in Fig.~\ref{fig:results_1}(b). A strong redshift at negative delays is followed by an equally large blueshift at positive pump-probe delays. During its propagation through the pump-affected region in the fused silica sample the probe pulse reads out refractive index variations induced by the strong excitation pulse. As a result the central probe wavelength varies according to 
\begin{equation}
\Delta  \omega(\tau) =  \frac{\partial \Delta \phi (\tau)}{\partial \tau} = - \frac{2 \pi L }{\lambda_0} \frac{\partial n(I_{\textrm{pump}}(t))}{\partial t} \otimes I_{\textrm{probe}}(t, \tau)
\end{equation}
where $L$ defines the interaction length, $\lambda_0$ the initial central probe wavelength and $n$ denotes the refractive index of the sample, which is related to the pump laser intensity $I_{\textrm{pump}}(t)$ via the optical Kerr-effect. $I_{\textrm{probe}}(t, \tau)$ is the probe laser intensity. As before, $\tau$ denotes the pump-probe delay and $\otimes$ the linear convolution operator. Thus, the integrated variation of the central probe wavelength reveals the delay dependence of the phase shift $\Delta \phi(\tau)$ induced by the intense SWIR laser pulse [see Fig.~\ref{fig:results_1}(b)]. Since XPM is linear in the intensity of the long wavelength driver, the delay dependence of $\Delta \phi(\tau)$ is determined by $\Delta \phi(\tau) \propto I_{\textrm{pump}}(t) \otimes  I_{\textrm{probe}}(t, \tau)$. Hence, the XPM-signal as a direct signature of the quasi-instantaneous optical Kerr-effect allows the reconstruction of the temporal envelope of the pump laser pulse assuming a well-characterized NIR probe laser pulse. Furthermore, this measurement provides an internal clock in the experiment enabling a measurement of the absolute timing of the simultaneously acquired transmission data with respect to the pump-probe overlap. \\
The measured time-resolved transmission of the probe laser pulse through the region affected by the pump is presented in Fig.~\ref{fig:results_1}(c). A double-peaked structure is observed with a first peak appearing close to the pump-probe overlap followed by a second peak delayed by $\approx\,$\SI{100}{\femto\second}. After the second peak the absorption slowly decays towards its initial value. We attribute the first peak to a second order nonlinear interaction of the pump and probe laser pulses that leads to a reversible depletion of the probe laser pulse and that was recently described as two-beam coupling \cite{Ye2018}. As shown in Ref.~\cite{Carr2005} an ionized region of dense plasma in a bulk solid exhibits a nonzero $\chi^{(2)}$ in contrast to the isotropic case of the pristine fused silica sample. We assign the second peak to absorption in the generated electron-hole plasma. In order to isolate the plasma response that is naturally screened by the reversible two-beam coupling effects \cite{Lechuga2017}, we apply a Gaussian fit to the leading edge of the first peak and subtract it from the total absorption curve. After doings so, we obtain the isolated plasma absorption as shown in Fig.~\ref{fig:results_1}(d). This curve reveals relaxation dynamics that are approximated by a single exponential decay time as suggested by e. g. \cite{Martin1997, Grojo2010}, leading to the determination of a relaxation time of $\tau_{\textrm{rel}} =\,$\SI{110}{\femto\second}.
In order to further analyze the relaxation dynamics of the generated electron-hole plasma, we performed the time-resolved XPM and transmission measurements as a function of the SWIR excitation intensity. Results obtained for the maximum value of the plasma absorption and the fitted relaxation time are summarized in Fig.~\ref{fig:results_3}(a) + (b).
Due to the highly nonlinear excitation process the maximum value of the absorption associated with the formation of the electron-hole plasma monotonously increases as a function of excitation intensity in a nonlinear fashion [see Fig.~\ref{fig:results_3}(a)]. Surprisingly, the fitted relaxation time displays a similar behavior. While only intensity-independent relaxation times have been observed in SiO$_2$ up to now \cite{Martin1997}, we detect an extremely fast relaxation with a characteristic time of $\approx\,$\SI{50}{\femto\second} at intensities far below the damage threshold. Once the intensity of the SWIR excitation pulse approaches the threshold for irreversible modification of the fused silica target [see colored areas in Fig.~\ref{fig:results_3}(a) + (b)] the relaxation time increases. Still, further increase of the excitation intensity to values clearly above the damage threshold leads to the observation of even longer decay times. \\
We attribute this observation to a detrapping mechanism initiated by the energy stored in the free carrier bath. This mechanism becomes more important when the laser energy is coupled into the material by a laser field of longer wavelength. Generally, a trapped carrier can be detrapped if sufficient energy is coupled into the system. At the same time the carrier can suffer changes in the ability to be trapped due to vibrational effects on the deformation potential. As shown by many studies, thermal detrapping occurs in various materials (see e.g. \cite{Zhang1998, Said2016} and references therein). Here, energy is coupled from the electron-hole plasma to the lattice by electron-phonon coupling leading to vibrations of the lattice. Once the vibrational energy is sufficient, a STE can be converted back into a quasi-free electron hole pair effectively prolonging the measured relaxation time of the laser-induced plasma. 
The energy stored in the electron-hole plasma can be estimated by 
\begin{equation} \label{eq:efcb}
E_{\textrm{e}^{-} \textrm{h}^{+}} \approx \rho(I) \times V \times \left[ E_{\textrm{gap}} + U_{\textrm{p}}(I) \right]
\end{equation}
with plasma density $\rho(I)$, volume $V$, bandgap energy $E_{\textrm{gap}}$ and ponderomotive energy $U_{\textrm{p}}$. The ponderomotive energy, i. e. the cycle-averaged electron quiver energy is defined through $U_p = e^2 E_{0}^{2}/4 \pi m^{\ast} \omega^2$. Here, $e$ is the elementary charge, $E_0$ is the amplitude of the laser electric field, $m^{\ast}$ denotes the reduced electron mass in the conduction band ($m^{\ast} = 0.635\,m_e$ for fused silica) and $\omega$ is the central laser frequency of the driving field. Equation~\ref{eq:efcb} can be evaluated by solving a single rate equation for the temporal evolution of the conduction band electron density given by \cite{Stuart1996, Mero2005b, Gulley2012}
\begin{equation} \label{eq:sre}
\frac{\partial \rho(t)}{\partial t} = \Gamma_{\textrm{SFI}} + \Gamma_{\textrm{IMP}} - \frac{\rho(t)}{\tau_{\textrm{rel}}} .
\end{equation}
Here, $\Gamma_{\textrm{SFI}}$ denotes the generation of quasi-free carriers by strong-field ionization, $\Gamma_{\textrm{IMP}}$ describes the variation of the conduction band electron density induced by electron-impact ionization, while the relaxation of quasi-free carriers from the conduction band is approximated by a single exponential decay time $\tau_{\textrm{rel}}$ that, amongst other processes, takes STE formation into account. Numerical solutions of Eq.~\ref{eq:efcb} as a function of the pump laser intensity are presented in Fig.~\ref{fig:results_3}(c) using the strong-field ionization rate provided by Keldysh \cite{Keldysh1965}, a modified Drude model for electron-impact ionization as presented in \cite{Jupe2009} and common material parameters for fused silica ($E_g = \,$\SI{7.5}{\electronvolt}, $\tau_{\textrm{rel}} = \,$\SI{220}{\femto\second}). The calculations are carried out for a NIR and a SWIR pump laser wavelength and a pulse duration corresponding to 20 optical cycles. Clearly, the energy coupled into the electron-hole plasma is much larger in the \SI{2100}{\nano\meter} case above a certain threshold intensity [$\approx \,$\SI{15}{\tera\watt\per\centi\meter\squared} in Fig.~\ref{fig:results_3}(c)]. Comparing this observation to the simulation of the relative contribution of IMP to the total generated conduction band electron density [see Fig.~\ref{fig:results_3}(d)] indicates that this threshold intensity corresponds to the onset of IMP playing a significant role in the formation of the electron-hole plasma. Since no propagation effects are included in the simulation the intensity values in the simulation are significantly above the experimental values. Naturally, the nonlinear effects like self-focusing lead to an increase in the laser pulse intensity inside the material, and hence a decrease of the damage threshold and an earlier onset of plasma formation \cite{Winkler2006}. \\
The differences between the NIR and SWIR scenario found in the simulations indicate that the signature of vibrational detrapping will be more pronounced in the case of a long wavelength driving field due to the higher average kinetic energy of the excited electrons. The fact that IMP plays a major role in the formation of the observed electron-hole plasma is furthermore supported by the delay of the maximum plasma density with respect to the overlap of the pump and probe pulse. Hence, the observation of an intensity-dependent relaxation time requires sufficient vibrational energy coupled into the lattice to detrap trapped carriers while a significant contribution of IMP to the total generated density of quasi-free carriers further assists this vibrational detrapping mechanism. \\ 
In conclusion, time-resolved transmission and XPM measurements have been performed on amorphous fused silica samples excited by femtosecond SWIR laser pulses, and a strong dependence of the relaxation time of excited conduction band electrons on the present electric field strength has been observed. We attribute this dependence to strongly increased energy coupling into the free carrier bath at the SWIR pump wavelengths used in the experiment leading to vibrational detrapping. This interpretation is in good agreement with intensity-independent trapping times, that were previously reported in the UV and in the VIS/NIR spectral domain, since at these wavelengths the ponderomotive energy and the influence of IMP are not sufficient to induce vibrational detrapping. \\
Our results extend the present literature on time-resolved studies of laser-matter interaction into the SWIR spectral domain and provide a novel method to isolate the persistent plasma response from transient, reversible effects based on the simultaneous detection of an XPM signal in bulk solids. Our findings provide a step towards understanding the mechanisms underlying laser micromachining with long wavelength driving fields and constitute the first observation of the influence of lattice vibrations on the trapping efficiency.
\section*{Acknowledgements}
The authors would like to acknowledge the financial support of the German Research Foundation DFG (grant no. ME 4427/1-1 \& ME 4427/1-2). \\

\section*{References}
\bibliography{Trapping_Dynamics_bib}% Produces the bibliography via BibTeX.

\end{document}